\documentclass[12pt,a4paper]{article}

\usepackage{cite}

\usepackage[latin1]{inputenc}
\usepackage{lineno}  
\usepackage{graphicx}  
\usepackage{xspace} 
\usepackage{booktabs} 
\usepackage{multirow}
\usepackage{color}
\usepackage{colortbl}
\usepackage{amsmath} 
\usepackage{ifthen} 
\usepackage{subfigure}
\usepackage{lscape} 

\usepackage{indentfirst}

\newboolean{pdflatex}
\setboolean{pdflatex}{true} 

\newboolean{articletitles}
\setboolean{articletitles}{true} 

\newboolean{uprightparticles}
\setboolean{uprightparticles}{false} 

\usepackage{amssymb}
\usepackage{amsfonts}
\usepackage{upgreek} 

\usepackage{ifpdf}

\usepackage{hyperref}    
\usepackage[all]{hypcap} 

\def\ux85 {UX85\xspace}

\ifthenelse{\boolean{uprightparticles}}%
{
 
 \def\Pgamma      {\ensuremath{\upgamma}\xspace}

 \def\Ppi         {\ensuremath{\uppi}\xspace}

 \def\PDelta      {\ensuremath{\Delta}\xspace}                 
 \def\PXi      {\ensuremath{\Xi}\xspace}                 
 \def\PLambda      {\ensuremath{\Lambda}\xspace}                 
 \def\PSigma      {\ensuremath{\Sigma}\xspace}                 
 \def\POmega      {\ensuremath{\Omega}\xspace}                 
 \def\PUpsilon      {\ensuremath{\Upsilon}\xspace}                 
 
 \def\PB      {\ensuremath{\mathrm{B}}\xspace}                 
                  
 \def\PD      {\ensuremath{\mathrm{D}}\xspace}

 \def\PK      {\ensuremath{\mathrm{K}}\xspace}

 \def\Pb      {\ensuremath{\mathrm{b}}\xspace}                 
 \def\Pc      {\ensuremath{\mathrm{c}}\xspace}

 \def\Pi      {\ensuremath{\mathrm{i}}\xspace}

 \def\Pu      {\ensuremath{\mathrm{u}}\xspace}

}
{
 
 \def\Pgamma      {\ensuremath{\gamma}\xspace}

 \def\Ppi         {\ensuremath{\pi}\xspace}

 \mathchardef\PDelta="7101
 \mathchardef\PXi="7104
 \mathchardef\PLambda="7103
 \mathchardef\PSigma="7106
 \mathchardef\POmega="710A
 \mathchardef\PUpsilon="7107
                  
 \def\PB      {\ensuremath{B}\xspace}                 
                  
 \def\PD      {\ensuremath{D}\xspace}

 \def\PK      {\ensuremath{K}\xspace}

 \def\Pb      {\ensuremath{b}\xspace}                 
 \def\Pc      {\ensuremath{c}\xspace}

 \def\Pi      {\ensuremath{i}\xspace}

 \def\Pu      {\ensuremath{u}\xspace}

}

\def\g      {\ensuremath{\Pgamma}\xspace}

\def\uquark    {\ensuremath{\Pu}\xspace}

\def\cquark    {\ensuremath{\Pc}\xspace}

\def\bquark    {\ensuremath{\Pb}\xspace}

\def\pion  {\ensuremath{\Ppi}\xspace}

\def\pip   {\ensuremath{\pion^+}\xspace}
\def\pim   {\ensuremath{\pion^-}\xspace}

\def\pipm  {\ensuremath{\pion^\pm}\xspace}

\def\kaon  {\ensuremath{\PK}\xspace}
  \def\Kbar  {\kern 0.2em\overline{\kern -0.2em \PK}{}\xspace}

\def\Kz    {\ensuremath{\kaon^0}\xspace}
\def\Kzb   {\ensuremath{\Kbar^0}\xspace}
\def\KzKzb {\ensuremath{\Kz \kern -0.16em \Kzb}\xspace}
\def\Kp    {\ensuremath{\kaon^+}\xspace}
\def\Km    {\ensuremath{\kaon^-}\xspace}
\def\Kpm   {\ensuremath{\kaon^\pm}\xspace}

\def\KpKm  {\ensuremath{\Kp \kern -0.16em \Km}\xspace}
\def\KS    {\ensuremath{\kaon^0_{\rm\scriptscriptstyle S}}\xspace} 
 
\def\Kstarz  {\ensuremath{\kaon^{*0}}\xspace}

\def\Kstarpm {\ensuremath{\kaon^{*\pm}}\xspace}

  \def\Dbar    {\kern 0.2em\overline{\kern -0.2em \PD}{}\xspace}
\def\D       {\ensuremath{\PD}\xspace}

\def\Dz      {\ensuremath{\D^0}\xspace}
\def\Dzb     {\ensuremath{\Dbar^0}\xspace}
\def\DzDzb   {\ensuremath{\Dz {\kern -0.16em \Dzb}}\xspace}
\def\Dp      {\ensuremath{\D^+}\xspace}
\def\Dm      {\ensuremath{\D^-}\xspace}

\def\DpDm    {\ensuremath{\Dp {\kern -0.16em \Dm}}\xspace}

\def\Dstarz  {\ensuremath{\D^{*0}}\xspace}

\def\B       {\ensuremath{\PB}\xspace}
  \def\Bbar    {\kern 0.18em\overline{\kern -0.18em \PB}{}\xspace}

\def\Bpm     {\ensuremath{\B^\pm}\xspace}

\def\Bd      {\ensuremath{\B^0}\xspace}

  \def\Y#1S{\ensuremath{\PUpsilon{(#1S)}}\xspace}

\newcommand{\decay}[2]{\ensuremath{#1\!\to #2}\xspace}         

\def\to                 {\ensuremath{\rightarrow}\xspace}

\def\CP                {\ensuremath{C\!P}\xspace}

\def\AT#1     {\ensuremath{A_{\mathrm{T}}^{#1}}\xspace}           

\def\C#1      {\ensuremath{\mathcal{C}_{#1}}\xspace}                       
\def\Cp#1     {\ensuremath{\mathcal{C}_{#1}^{'}}\xspace}                    
\def\Ceff#1   {\ensuremath{\mathcal{C}_{#1}^{\mathrm{(eff)}}}\xspace}        
\def\Cpeff#1  {\ensuremath{\mathcal{C}_{#1}^{'\mathrm{(eff)}}}\xspace}       
\def\Ope#1    {\ensuremath{\mathcal{O}_{#1}}\xspace}                       
\def\Opep#1   {\ensuremath{\mathcal{O}_{#1}^{'}}\xspace}                    

\newcommand{\tev}{\ensuremath{\mathrm{\,Te\kern -0.1em V}}\xspace}
\newcommand{\gev}{\ensuremath{\mathrm{\,Ge\kern -0.1em V}}\xspace}
\newcommand{\mev}{\ensuremath{\mathrm{\,Me\kern -0.1em V}}\xspace}
\newcommand{\kev}{\ensuremath{\mathrm{\,ke\kern -0.1em V}}\xspace}
\newcommand{\ev}{\ensuremath{\mathrm{\,e\kern -0.1em V}}\xspace}
\newcommand{\gevc}{\ensuremath{{\mathrm{\,Ge\kern -0.1em V\!/}c}}\xspace}
\newcommand{\mevc}{\ensuremath{{\mathrm{\,Me\kern -0.1em V\!/}c}}\xspace}
\newcommand{\gevcc}{\ensuremath{{\mathrm{\,Ge\kern -0.1em V\!/}c^2}}\xspace}
\newcommand{\gevgevcccc}{\ensuremath{{\mathrm{\,Ge\kern -0.1em V^2\!/}c^4}}\xspace}
\newcommand{\mevcc}{\ensuremath{{\mathrm{\,Me\kern -0.1em V\!/}c^2}}\xspace}

\def\invfb   {\ensuremath{\mbox{\,fb}^{-1}}\xspace}

\def\gsim{{~\raise.15em\hbox{$>$}\kern-.85em
          \lower.35em\hbox{$\sim$}~}\xspace}
\def\lsim{{~\raise.15em\hbox{$<$}\kern-.85em
          \lower.35em\hbox{$\sim$}~}\xspace}

\def\degrees{\ensuremath{^{\circ}}\xspace}

\def\tell1  {TELL1\xspace}
\def\ukl1   {UKL1\xspace}

\newcommand{\ie}{\mbox{\itshape i.e.}}

\newcolumntype{e}{@{ $\pm$ }l}         
\newcolumntype{R}{>{$}r<{$}}           
\newcolumntype{L}{>{$}l<{$}}           
\newcolumntype{C}{>{$}c<{$}}           
\newcolumntype{E}{@{ $\pm$ }>{$}l<{$}} 

\usepackage{wasysym}     

\usepackage{mciteplus}

\newcommand{\re}[2][()] {\ifthenelse{\equal{#1}{()}}{{\ensuremath{{\rm \, Re}}\left(#2\right)}}
                                                    {{\ensuremath{{\rm \, Re}}\left[#2\right]}}}
\newcommand{\im}[2][()] {\ifthenelse{\equal{#1}{()}}{{\ensuremath{{\rm \, Im}}\left(#2\right)}}
                                                    {{\ensuremath{{\rm \, Im}}\left[#2\right]}}}

\newcommand{\afb}{A_{\bar{D}}}
\newcommand{\af} {\vphantom{\afb}A_D}

\newcommand{\inbars}[1]{\left|#1\right|}
\newcommand{\inbarssq}[1]{\left|#1\right|^2}

\newcommand{\dpmcpket}{\left|     {D}_\pm^{\mathrm{CP}}      \right\rangle}

\newcommand{\dzket}   {\left|       \vphantom{\bar{D}^0}     {D}^0      \right\rangle}
\newcommand{\dzbket}  {\left|                            \bar{D}^0      \right\rangle}

\newcommand{\mrow}[2]   {\multirow{#1}{*}{#2}}

\newcommand{\mcolr}[2]  {\multicolumn{#1}{r}{#2}}

\newcommand{\bdk}    {\decay{\Bpm}{\D\Kpm}}
\newcommand{\bdpi}   {\decay{\Bpm}{\D\pipm}}
\newcommand{\bzdkst} {\decay{\Bd} {\D\Kstarz}}

\newcommand{\bdkst}  {\decay{\Bpm}{\D\Kstarpm}}
\newcommand{\bdstk}  {\decay{\Bpm}{\Dstarz                   \Kpm}}

\newcommand\uncheckbox{\makebox[0pt][l]{$\square$}\raisebox{.15ex}{\hspace{0.1em}\hphantom{$\checkmark$}}}

\begin{document}

\begin{titlepage}

\vspace*{-1.5cm}

\hspace*{-0.5cm}

\vspace*{4.0cm}

{\bf\boldmath\huge
\begin{center}
Study of the sensitivity to CKM angle $\gamma$ under simultaneous determination from multiple
$B$ meson decay modes
\end{center}
}

\vspace*{2.0cm}

\begin{center}
	J. Garra Tic\'o$^1$, V. Gibson$^2$, S. C. Haines$^1$, C. R. Jones$^2$, M. Kenzie$^{1,3}$, G. Lovell$^2$
\bigskip\\
{\it\footnotesize
$ ^1$Formerly Cavendish Laboratory, University of Cambridge, Cambridge CB3 0HE, United Kingdom\\
$ ^2$Cavendish Laboratory, University of Cambridge, Cambridge CB3 0HE, United Kingdom\\
$ ^3$Department of Physics, University of Warwick, Coventry CV4 7AL, United Kingdom\\
}
\end{center}

\vspace{\fill}

\begin{abstract}
  Several methods exist to measure \CP violation observables related to
  the CKM unitarity triangle angle $\gamma$ using $B$ meson decays.
  These observables are different for every $B$ meson decay considered,
  although the information they contain on $\gamma$ is encoded in a similar way for all of them.
  This paper describes a strategy for a simultaneous measurement of $\gamma$ using several $B$ meson decays that takes into account possible correlations between them based on the methodologies described in \cite{1804.05597}.
  Sensitivity studies demonstrate that the simultaneous analysis of several $B$ meson decay modes results in smaller uncertainties and improved statistical
  behaviour compared to a combination of standalone measurements.
\end{abstract}

\vspace*{2.0cm}
\vspace{\fill}

\end{titlepage}

\thispagestyle{empty}  

\newpage
\setcounter{page}{2}
\mbox{~}



\section{Introduction}
\label{sc.intro}

The angle $\gamma \equiv \arg\left( - \frac{V_{ud}V^\star_{ub}}{V_{cd}V^\star_{cb}} \right)$ of the Cabibbo-Kobayashi-Maskawa (CKM) unitarity triangle
can be measured using tree-level $B$ meson decays that involve interference between $\bquark \to
\uquark$ and $\bquark \to \cquark$ quark transitions.
Time-integrated measurements can be made to measure $\gamma$ using decays of the type $B \to D X$,
where $D$ represents an admixture of the flavour eigenstates
\Dz\ and \Dzb and $X$ a final state containing one or more kaons or pions.
Examples include
$\Bpm \to \D^{(*)} K^{(*)\pm}$, $\Bd \to \D \Kstarz$ and
$\Bpm \to \D^{(*)} \pipm$ decays.
Alongside this, a time-dependent approach can be employed to measure $\gamma$ from
decays such as $B^0_s \to D^-_s K^+$. More details on the extraction
of the CKM angle $\gamma$ can be found in Refs.~\cite{PDG2020,HFLAV18,Gershon:2016fda}.

Several methods can be utilised to measure different \CP violation observables in these decays~\cite{glw1,glw2,ads3,ads4,ggsz,ggsz2,td}. The measurements are then typically used to place tree-level constraints on $\gamma$ without the need for any theoretical input.
The current world average value of $\gamma = (71.1^{+4.6}_{-5.3})^{\circ}$~\cite{PDG2018,HFLAV18}
is dominated by measurements from the LHCb experiment~\cite{LHCb:GammaCombo2016,LHCb:GammaCombo2018}.

This paper presents results for a simultaneous approach to constrain $\gamma$ from multiple $\B$ meson decays~\cite{1804.05597},
allowing for the treatment of experimental candidates reconstructed under different decay hypotheses
and for the straightforward determination of correlations between systematic uncertainties.
The technique employs a reduced set of \CP violation parameters and is applicable to
all possible measurement approaches. The results presented in this paper make a comparison between the simultaneous method and the traditional approach of fitting for each decay mode independently.

\section{Simultaneous approach for time-integrated measurements with {\boldmath $B \to D X$} decays}
\label{sc.approach}

\newcommand{\dsket}[1]{\left| \vphantom{\bar{D}^0}     {D}_{#1} \right\rangle}

\newcommand{\defaf}  {\left\langle f \left| \mathcal{H} \vphantom{D^0} \right| \! \vphantom{\bar{D}^0}     {D}^0     \right\rangle}
\newcommand{\defafb} {\left\langle f \left| \mathcal{H} \vphantom{D^0} \right| \!                      \bar{D}^0     \right\rangle}
\newcommand{\defap}  {\left\langle f \left| \mathcal{H} \vphantom{D^0} \right| \! \vphantom{\bar{D}^0}     {D}_+     \right\rangle}
\newcommand{\defam}  {\left\langle f \left| \mathcal{H} \vphantom{D^0} \right| \! \vphantom{\bar{D}^0}     {D}_-     \right\rangle}
\newcommand{\defapm} {\left\langle f \left| \mathcal{H} \vphantom{D^0} \right| \! \vphantom{\bar{D}^0}     {D}_\pm   \right\rangle}
\newcommand{\defapmc}{\left\langle f \left| \mathcal{H} \vphantom{D^0} \right| \! \vphantom{\bar{D}^0}     {D}_\pm^m \right\rangle}

\newcommand{\zpm} {z_\pm}
\newcommand{\zpmc}{z_\pm^m}

\subsection{Admixture coefficients {\boldmath $\zpmc$}}
\label{sc.formalism}

As described above, several $B$ meson decays produce admixtures of neutral $D$ mesons that involve $\gamma$.
In this paper, $\dzket$ and $\dzbket$ are used to denote the flavour eigenstates of neutral $D$ mesons;
$\dsket{+}$ represents the $D$ meson produced in $B^+$ or $B^0$ meson decays
and $\dsket{-}$ the
$D$ meson produced in $B^-$ or $\bar{B}^0$ decays.

In general, one can write
\begin{equation}
  \begin{aligned}
    \left| D_-^m \right\rangle &\sim \left|     {D}^0 \right\rangle + z_-^m \left| \bar{D}^0 \right\rangle \\
    \left| D_+^m \right\rangle &\sim \left| \bar{D}^0 \right\rangle + z_+^m \left|     {D}^0 \right\rangle
  \end{aligned}
  \Rightarrow
  \left\{
  \begin{aligned}
    A_-^m &\sim \af  + z_-^m \afb \\
    A_+^m &\sim \afb + z_+^m \af ,
  \end{aligned}
  \right.
\end{equation}
where
$m$ denotes the $B$ decay mode under consideration.
The amplitudes $\af = \defaf$, $\afb = \defafb$ and $A_\pm^m = \defapmc$
define the $D$ meson decay to a final state $f$.
The complex coefficients $\zpmc$ are specific to each $B$ decay, and are typically expressed in either Cartesian $(x_\pm^m, y_\pm^m)$ or polar $(r_m, \delta_m, \gamma)$ coordinates as
\begin{equation}
  z_\pm^m = x_\pm^m + i\,y_\pm^m = r_m\, e^{i \delta_m} \, e^{\pm i \gamma},
\end{equation}
where all parameters with subscript or superscript $m$ are specific to a particular $B$ meson decay.
It is apparent that $r_m$ and $\delta_m$ represent the ratio of amplitude magnitudes and their strong phase difference.
Using the definition
\begin{equation} \label{eq.zc}
  z_m = r_m \, e^{i \delta_m},
\end{equation}
leads to
\begin{equation}
  z_\pm^m = z_m \, e^{\pm i \gamma},
\end{equation}
and reveals an invariant for each $B$ meson decay,
\begin{equation}
  \frac{z_+^m}{z_-^m} = e^{2 i \gamma}
  \quad \Rightarrow \quad
  \gamma = \frac{1}{2} \arg\left(\frac{z_+^m}{z_-^m}\right).
\end{equation}
If both $\zpmc$ coefficients are multiplied by any complex coefficient $\xi$, the result will contain
exactly the same information on $\gamma$. 
In particular, it is always possible to relate the $\zpmc$ coefficients for channel $m$
to those for $\Bpm \to \D K^{\pm}$ decays, denoted $z_\pm$:
\begin{equation} \label{eq.xicdefn}
  z_\pm^m = \xi_m \, z_\pm,
\end{equation}
where
\begin{equation} \label{eq.xic}
  \xi_m = \frac{z_m}{z_{\mathrm{DK}}}.
\end{equation}
By definition, from equations (\ref{eq.zc}) and (\ref{eq.xic}), the $\xi_m$ coefficients do not depend
on $\gamma$ and can, therefore, be considered as nuisance parameters.

When global averages are extracted for $\gamma$~\cite{PDG2020}, many of the different input measurements depend on the hadronic unknowns, $r_m$ and $\delta_m$, as well as $\gamma$. Subsequently, it is not valid to simply average the various measurements of $\gamma$. Instead, individual measurements extract values of $z_\pm^m$ which are used as inputs to a global combination.
Thus, using the $\xi_m$ coefficients to perform a simultaneous fit for the Cartesian parameters in $N$ distinct $B$ meson decay modes,
reduces the number of independent parameters in the fit from $4\,N$ to $4 + 2(N-1)$: $4$ parameters for $\Bpm \to \D K^{\pm}$, and $2$ for each of the other decays.
Although this is only one more parameter than a
simultaneous fit for the $1 +2N$ polar coordinates ($\gamma$, $r_m$, $\delta_m$), it has the advantage that the real
and imaginary components of $\zpm$ and $\xi_m$ are expected to exhibit Gaussian behaviour and can be used in conjunction with other orthogonal measurements in a global combination.

\subsection{The {\boldmath $\eta$} function}

In order to simplify later notation, it is useful to define the $\eta$ function as
\begin{equation}
  \eta\left( a, b, \kappa \right) = \left| a \right|^2 + \left| b \right|^2 + 2 \kappa \re{ a^\star b },
\end{equation}
where $a,b \in \mathbb{C}$ and $\kappa \in \mathbb{R}$.
This function is symmetric with respect to the exchange of $a$ and $b$ ($\eta(a, b, \kappa) = \eta(b, a, \kappa)$)
and scales as $\eta(a, b, \kappa) = \inbarssq{a} \eta\left(1,\frac{b}{a},\kappa\right)$.
The $\kappa$ coefficient, commonly known as the \textit{coherence factor}, indicates the fraction of coherent sum that contributes to $\eta$,
\begin{equation}
  \eta\left( a, b, \kappa \right) = \kappa \left| a + b \right|^2 + ( 1 - \kappa ) \left( \left| a \right|^2 + \left| b \right|^2 \right).
\end{equation}
In this paper, when the coherence factor argument is omitted, it should be assumed that it is implicit and,
if one of the complex arguments is omitted, it should be assumed that it is $1$
(for example, $ \eta\left( a \right) \equiv \eta\left( a, 1, \kappa \right)$).

\subsection{Signal amplitude}

The probability distribution function for $B$ mesons to decay via a particular decay mode $m$ is proportional to the squared amplitude $\left| A_\pm^m \right|^2$ integrated over the phase space of the final state particles in the $B$ decay,
\begin{equation}
  p_\pm^m \sim \int d\mathcal{P}_B \left| A_\pm^m \right|^2.
\end{equation}

For a specific $B$ decay mode, defining $A_c$ as the decay amplitude corresponding to a $b \rightarrow c$ transition and $A_u \, e^{\pm i \gamma}$ as
the decay amplitude corresponding to a $b \rightarrow u$ transition leads to
\begin{align}
  A_- &\sim A_c \af  + A_u e^{-i \gamma} \afb, \\
  A_+ &\sim A_c \afb + A_u e^{+i \gamma} \af .
\end{align}

For a 2-body $B$ meson decay, such as $B^\pm \to D K^\pm$, the amplitudes $A_c$ and $A_u$ are
simply the magnitude of the transition amplitude integrated over 
the phase space, and one can write
\begin{align}
  A_- &\sim \af  + z_- \afb, \label{eq.am} \\
  A_+ &\sim \afb + z_+ \af , \label{eq.ap}
\end{align}
where
\begin{equation}
  z_\pm = \frac{A_u}{A_c} e^{\pm i \gamma}.
\end{equation}
This implies that, for 2-body decays,
\begin{align}
  p_- &\sim \eta( \af , z_-\, \afb ), \\
  p_+ &\sim \eta( \afb, z_+\, \af  ).
\end{align}

In the case of a multi-body $B$ meson decay with $3$ or more particles in the final state,
such as $B^0 \to D K \pi$,
the amplitude $\left| A_\pm^m \right|^2$ may be integrated over a reduced part of the $B$ decay phase space, for example around the $K^{*0} (892)$ resonance for $B^0 \to D K \pi$.

By squaring the modulus of expressions (\ref{eq.am}) and (\ref{eq.ap}) and defining
\begin{align}
  N_{\alpha\beta} &= \int d\mathcal{P}_B A^\star_\alpha A_\beta, \\
  X_{\alpha\beta} &= \frac{N_{\alpha\beta}}{\sqrt{N_{\alpha\alpha} N_{\beta\beta}}},
\end{align}
one can write
\begin{align}
  p_- 
      &\sim        \inbarssq{\af}  + \frac{N_{uu}}{N_{cc}} \inbarssq{\afb} +
            2 \inbars{X_{cu}} \re{\sqrt{\frac{N_{uu}}{N_{cc}}}\,\frac{X_{cu}}{\inbars{X_{cu}}} e^{-i\gamma} \af^\star \afb }, \\
  p_+ 
      &\sim        \inbarssq{\afb} + \frac{N_{uu}}{N_{cc}} \inbarssq{\af}  +
            2 \inbars{X_{cu}} \re{\sqrt{\frac{N_{uu}}{N_{cc}}}\,\frac{X_{cu}}{\inbars{X_{cu}}} e^{+i\gamma} \afb^\star \af }.
\end{align}

It should be noted that $\left| X_{\alpha\beta} \right| \leq 1$, because of the Cauchy-Schwarz inequality.
Defining
\begin{align}
  \kappa      &= \inbars{X_{cu}}, \\
  r           &= \sqrt{\frac{N_{uu}}{N_{cc}}}, \\
  e^{i\delta} &= \frac{X_{cu}}{\inbars{X_{cu}}}, \\
  z           &= r\, e^{i\delta}, \\
  z_\pm       &= z\, e^{\pm i \gamma},
\end{align}
the signal amplitude probability distribution for $B$ meson decay mode $m$ can then be expressed as
\begin{align}
  p^m_- &\sim \eta( \af , \xi_m \, z_- \, \afb, \kappa_m ) =
  \inbarssq{\af}  + \inbarssq{\xi_m \, z_-} \inbarssq{\afb} + 2 \, \kappa_m \, \re{ \xi_m \, z_- \af ^\star \afb }, \label{eq.pcm} \\
  p^m_+ &\sim \eta( \afb, \xi_m \, z_+ \, \af , \kappa_m ) =
  \inbarssq{\afb} + \inbarssq{\xi_m \, z_+} \inbarssq{\af}  + 2 \, \kappa_m \, \re{ \xi_m \, z_+ \afb^\star \af  }. \label{eq.pcp}
\end{align}

These expressions describe the physics of the neutral $D$ meson admixture that leads to the
different \CP observables used to measure $\gamma$, but they are not specific to any measurement method.

\newcommand{\witheta}{\boolean{true}}
\newcommand{\Kst}{\ensuremath{K^\star}\xspace}
\newcommand{\Kstp}{\ensuremath{K^{\star+}}\xspace}
\newcommand{\Kstm}{\ensuremath{K^{\star-}}\xspace}
\newcommand{\Kstpm}{\ensuremath{K^{\star\pm}}\xspace}

\section{Specific formalism for established methodologies}
\label{sc.eqs}

\subsection{Decays to CP eigenstates}

This method (commonly referred to as the GLW method~\cite{glw1,glw2}) uses two sets of final states from the $D$ meson decay.
The first are those states that are accessible from only one of the $D$ meson flavour eigenstates,
either $\dzket$ or $\dzbket$, such that
$A^{    {D}}_{    {f}} = \left\langle                        f_{    {D}} | \mathcal{H} |                       {D}^0 \right\rangle$,
$A^{\bar{D}}_{\bar{f}} = \left\langle                        f_{\bar{D}} | \mathcal{H} |                   \bar{D}^0 \right\rangle$, and
$                        \left\langle                        f_{\bar{D}} | \mathcal{H} | \vphantom{\bar{D}^0}  {D}^0 \right\rangle =
                         \left\langle \vphantom{f_{\bar{D}}} f_{    {D}} | \mathcal{H} |                   \bar{D}^0 \right\rangle = 0$.
The second are those states that are accessible from one of the $CP$ eigenstates $\dpmcpket$, such that
$A^{\mathrm{CP}}_\pm = \left\langle f_\pm | \mathcal{H} | D^\mathrm{CP}_\pm \right\rangle$
and $\left\langle f_\mp | \mathcal{H} | D^\mathrm{CP}_\pm \right\rangle = 0$.

The observables of interest for a given $B$ decay mode $m$ are ratios and asymmetries
that can be used to constrain $\zpm$ and $\xi_{m}$.
For example, for $\Bpm \to \D \pipm$ decays,
\begin{align}
  R_{\mathrm{CP}}^{\pm \, D\pi}
  &=
  \frac{\Gamma\left( B^- \rightarrow D_\pm^\mathrm{CP} \pi^- \right) +
        \Gamma\left( B^+ \rightarrow D_\pm^\mathrm{CP} \pi^+ \right)}
       {\Gamma\left( B^- \rightarrow     {D}^0         \pi^- \right) +
        \Gamma\left( B^+ \rightarrow \bar{D}^0         \pi^+ \right)} 
  =
  \frac{1}{2} \left| \frac{A^{\mathrm{CP}}_\pm}{A^D_f} \right|^2
  \ifthenelse{\witheta}
  {
    \frac{\eta\left( \pm \xi_{D\pi} \, z_- \right) + \eta\left( \pm \xi_{D\pi} \, z_+ \right)}
         {2},
  }
  {
    \frac{\left| 1 \pm \xi_{D\pi} \, z_- \right|^2 + \left| 1 \pm \xi_{D\pi} \, z_+ \right|^2}
         {2},
  }
  \\
  A_{\mathrm{CP}}^{\pm \, D\pi}
  &=
  \frac{\Gamma\left( B^- \rightarrow D_\pm^\mathrm{CP} \pi^- \right) -
        \Gamma\left( B^+ \rightarrow D_\pm^\mathrm{CP} \pi^+ \right)}
       {\Gamma\left( B^- \rightarrow D_\pm^\mathrm{CP} \pi^- \right) +
        \Gamma\left( B^+ \rightarrow D_\pm^\mathrm{CP} \pi^+ \right)} 
  =
  \ifthenelse{\witheta}
  {
    \frac{\eta\left( \pm \xi_{D\pi} \, z_- \right) - \eta\left( \pm \xi_{D\pi} \, z_+ \right)}
         {\eta\left( \pm \xi_{D\pi} \, z_- \right) + \eta\left( \pm \xi_{D\pi} \, z_+ \right)}.
  }
  {
    \frac{\left| 1 \pm \xi_{D\pi} \, z_- \right|^2 - \left| 1 \pm \xi_{D\pi} \, z_+ \right|^2}
         {\left| 1 \pm \xi_{D\pi} \, z_- \right|^2 + \left| 1 \pm \xi_{D\pi} \, z_+ \right|^2}.
  }
\end{align}

\subsection{Decays to Cabibbo-favored and doubly-Cabibbo-suppressed final states}

This method (commonly referred to as ADS method~\cite{ads3,ads4}) uses final states that are accessible from both neutral $D$ meson flavour eigenstates,
enhancing the possible $CP$ asymmetry by considering the interference between
a favoured $B$ meson decay followed by a doubly CKM-suppressed $D$ decay, and
a suppressed $B$ meson decay followed by a CKM-favoured $D$ decay.
For the example of $\Bpm \to \D \pipm$ decays,
with the convention $CP \dzket = \dzbket$, and assuming no direct $CP$ violation in the $D$ decay,
\begin{alignat}{2}
  \Gamma^\pm_\mathrm{fav}
  &=
  \Gamma\left( B^\pm \rightarrow D_\mathrm{fav} \, \pi^\pm \right)
  =
  \ifthenelse{\witheta}
  {
    \left| A^\pm_{\tilde{D}} A^D_f \right|^2
    \eta\left( 1, \rho \, \xi_{D\pi} \, z_\pm \right)
  }
  {
    \left| A^\pm_{\tilde{D}} A^D_f \right|^2
    \left| 1 + \rho \, \xi_{D\pi} \, z_\pm \right|^2
  }
  ,
  \\
  \Gamma^\pm_\mathrm{sup}
  &=
  \Gamma\left( B^\pm \rightarrow D_\mathrm{sup} \, \pi^\pm \right)
  =
  \ifthenelse{\witheta}
  {
    \left| A^\pm_{\tilde{D}} A^D_f \right|^2
    \eta\left( \rho, \xi_{D\pi} \, z_\pm \right)
  }
  {
    \left| A^\pm_{\tilde{D}} A^D_f \right|^2
    \left| \rho + \xi_{D\pi} \, z_\pm \right|^2
  }
  ,
\end{alignat}
where $\rho =
  \frac{A^{\bar{D}}_{    {f}}}
       {A^{    {D}}_{    {f}}}$, and the subscripts ``sup" and ``fav" refer to the suppressed and favoured decay modes of
the produced $D$ meson, respectively.

The \CP observables of interest are
\begin{equation}
  R_{\mathrm{ADS}}^{\pm \, D\pi}
  =
  \frac{\Gamma^\pm_\mathrm{sup}}
       {\Gamma^\pm_\mathrm{fav}}
  =
  \ifthenelse{\witheta}
  {
    \frac{ \eta\left( \rho,         \xi_{D\pi} \, z_\pm \right) }
         { \eta\left( 1   , \rho \, \xi_{D\pi} \, z_\pm \right) }.
  }
  {
    \left|
      \frac{ \rho +         \xi_{D\pi} \, z_\pm }
           { 1    + \rho \, \xi_{D\pi} \, z_\pm }
    \right|^2.
  }
\end{equation}
Other ratios and asymmetries are also commonly used,
\begin{alignat}{2}
  R_{\mathrm{ADS}}^{D\pi}
  &=
  \frac{\Gamma^-_\mathrm{sup} + \Gamma^+_\mathrm{sup}}
       {\Gamma^-_\mathrm{fav} + \Gamma^+_\mathrm{fav}}
  &&=
  \ifthenelse{\witheta}
  {
    \frac{\eta\left( \rho, \xi_{D\pi} \, z_- \right) +
          \eta\left( \rho, \xi_{D\pi} \, z_+ \right)}
         {\eta\left(1, \rho \, \xi_{D\pi} \, z_- \right) +
          \eta\left(1, \rho \, \xi_{D\pi} \, z_+ \right)},
  }
  {
    \frac{\left| \rho +         \xi_{D\pi} \, z_- \right|^2 +
          \left| \rho +         \xi_{D\pi} \, z_+ \right|^2}
         {\left| 1    + \rho \, \xi_{D\pi} \, z_- \right|^2 +
          \left| 1    + \rho \, \xi_{D\pi} \, z_+ \right|^2},
  }
  \\
  A_{\mathrm{ADS}}^{\mathrm{sup} \, D\pi}
  &=
  \frac{\Gamma^-_\mathrm{sup} - \Gamma^+_\mathrm{sup}}
       {\Gamma^-_\mathrm{sup} + \Gamma^+_\mathrm{sup}}
  &&=
  \ifthenelse{\witheta}
  {
    \frac{\eta\left( \rho, \xi_{D\pi} \, z_- \right) -
          \eta\left( \rho, \xi_{D\pi} \, z_+ \right)}
         {\eta\left( \rho, \xi_{D\pi} \, z_- \right) +
          \eta\left( \rho, \xi_{D\pi} \, z_+ \right)},
  }
  {
    \frac{\left| \rho + \xi_{D\pi} \, z_- \right|^2 -
          \left| \rho + \xi_{D\pi} \, z_+ \right|^2}
         {\left| \rho + \xi_{D\pi} \, z_- \right|^2 +
          \left| \rho + \xi_{D\pi} \, z_+ \right|^2},
  }
  \\
  A_{\mathrm{ADS}}^{\mathrm{fav} \, D\pi}
  &=
  \frac{\Gamma^-_\mathrm{fav} - \Gamma^+_\mathrm{fav}}
       {\Gamma^-_\mathrm{fav} + \Gamma^+_\mathrm{fav}}
  &&=
  \ifthenelse{\witheta}
  {
    \frac{\eta\left( 1, \rho \, \xi_{D\pi} \, z_- \right) -
          \eta\left( 1, \rho \, \xi_{D\pi} \, z_+ \right)}
         {\eta\left( 1, \rho \, \xi_{D\pi} \, z_- \right) +
          \eta\left( 1, \rho \, \xi_{D\pi} \, z_+ \right)},
  }
  {
    \frac{\left| 1 + \rho \, \xi_{D\pi} \, z_- \right|^2 -
          \left| 1 + \rho \, \xi_{D\pi} \, z_+ \right|^2}
         {\left| 1 + \rho \, \xi_{D\pi} \, z_- \right|^2 +
          \left| 1 + \rho \, \xi_{D\pi} \, z_+ \right|^2},
  }
\end{alignat}
again showing the relationship between the observables and $\xi_{D\pi}$ and $\zpm$.

\subsection{Decays to multi-body self-conjugate final states}

This method (commonly referred to as the BPGGSZ method~\cite{ggsz,ggsz2,PhysRevD.70.072003})
uses $D$ meson decays to three or more final state particles that
can be accessed from both $\dzket$ or $\dzbket$ (e.g.~$D^0\to\KS\pip\pim$).
In contrast to the GLW or ADS approaches, this method does not involve intermediate
observables, and the goal is to fit for $\xi_{m}$ and $\zpm$ directly, using Equations
(\ref{eq.pcm}-\ref{eq.pcp}).

\subsection{Extension to time-dependent measurements}

The time evolution of the amplitude of the decay $B_s^0 \to D_s^{\mp} K^{\pm}$ is governed by the equation
\begin{equation}
  A^{B_s^0}_{D_s^\mp}(t)
  = A^{B_s^0}_{D_s^\mp} \left[ g_+(t) + \lambda_\pm g_-(t) \right],
\end{equation}
where $t$ is the $B_s^0$ meson lifetime and $g_\pm(t)$ are functions that describe the mixing of
the $B_s^0$ meson flavour eigenstates. The $\lambda_\pm$ parameters can be expressed as
\begin{equation}
  \lambda_\pm = \xi_m \, z_\pm \, e^{ \pm i \phi_q },
\end{equation}
where $\phi_q$ is the weak phase arising in the interference
between decay and mixing of the initial state neutral $B$-meson. In 
this case $m$ specifically denotes $B_s^0 \to D_s^{\mp} K^{\pm}$ and
$\phi_q = \phi_s$.

\newcommand{\lambdafsq}{\left|\lambda_f\right|^2}
\newcommand{\lambdamsq}{\left|\lambda_-\right|^2}

Existing measurements of $\gamma$ in decays with $B$ meson mixing introduce intermediate
observables instead of targeting the $z^m_\pm$ parameters themselves. In these cases, the squared
amplitude of the time-dependent probability distribution function is expressed as
\begin{equation}
  \small
  e^{-\Gamma t} \,
  \frac{1 + \lambdamsq}{2} 
  \left[\vphantom{\frac{1 - \lambdamsq }{1 + \lambdamsq}}\right.
    \cosh(x \Gamma t) +
    \underbrace{\frac{1 - \lambdamsq   }{1 + \lambdamsq}}_{\displaystyle C_f}                \cos (y \Gamma t)-
    \underbrace{\frac{-2 \re{\lambda_-}}{1 + \lambdamsq}}_{\displaystyle A_f^{\Delta\Gamma}} \sinh(x \Gamma t)-
    \underbrace{\frac{ 2 \im{\lambda_-}}{1 + \lambdamsq}}_{\displaystyle S_f}                \sin (y \Gamma t)
    \left.\vphantom{\frac{1 - \lambdamsq}{1 + \lambdamsq}}\right],
\end{equation}
where $x$ and $y$ are parameters describing the $B$ meson
mixing and $\Gamma$ is its decay rate; there is a similar expression for the conjugated amplitude. 
Although this approach~\cite{td} is observable based, it is possible to include the $C_f$, $A_f^{\Delta\Gamma}$ and $S_f$ parameters, and equivalent parameters for the conjugate decay, 
in a simultaneous fit that shares the $z_\pm$ parameters
with other $B$ decays and only introduces $\xi_m$ as a new coefficient.

\newcommand{\bl}{\hphantom{-}}

\newcommand{\DKstz}{{D\hspace{-1pt}K\hspace{-1pt}^{\star\hspace{-0.5pt}0}}}
\newcommand{\DKst} {{D\hspace{-1pt}K\hspace{-1pt}^{\star}}}
\newcommand{\DstK} {{D\hspace{-0.7pt}^\star\hspace{-1.5pt} K}}

\newcommand{\Dstz}{\ensuremath{D^{*0}}\xspace}
\newcommand{\Kstz}{\ensuremath{K^{*0}}\xspace}

\newcommand{\xmbdk}{\ensuremath{x_{-}^{\Dz\Kpm}}\xspace}
\newcommand{\ymbdk}{\ensuremath{y_{-}^{\Dz\Kpm}}\xspace}
\newcommand{\xpbdk}{\ensuremath{x_{+}^{\Dz\Kpm}}\xspace}
\newcommand{\ypbdk}{\ensuremath{y_{+}^{\Dz\Kpm}}\xspace}
\newcommand{\xixbdpi}{\ensuremath{x_{\xi}^{\Dz\pipm}}\xspace}
\newcommand{\xiybdpi}{\ensuremath{y_{\xi}^{\Dz\pipm}}\xspace}
\newcommand{\xmbdpi}{\ensuremath{x_{-}^{\Dz\pipm}}\xspace}
\newcommand{\ymbdpi}{\ensuremath{y_{-}^{\Dz\pipm}}\xspace}
\newcommand{\xpbdpi}{\ensuremath{x_{+}^{\Dz\pipm}}\xspace}
\newcommand{\ypbdpi}{\ensuremath{y_{+}^{\Dz\pipm}}\xspace}
\newcommand{\xixbdkstz}{\ensuremath{x_{\xi}^{\Dz\Kstz}}\xspace}
\newcommand{\xiybdkstz}{\ensuremath{y_{\xi}^{\Dz\Kstz}}\xspace}
\newcommand{\xmbdkstz}{\ensuremath{x_{-}^{\Dz\Kstz}}\xspace}
\newcommand{\ymbdkstz}{\ensuremath{y_{-}^{\Dz\Kstz}}\xspace}
\newcommand{\xpbdkstz}{\ensuremath{x_{+}^{\Dz\Kstz}}\xspace}
\newcommand{\ypbdkstz}{\ensuremath{y_{+}^{\Dz\Kstz}}\xspace}
\newcommand{\xixbdkst}{\ensuremath{x_{\xi}^{\Dz\Kstpm}}\xspace}
\newcommand{\xiybdkst}{\ensuremath{y_{\xi}^{\Dz\Kstpm}}\xspace}
\newcommand{\xmbdkst}{\ensuremath{x_{-}^{\Dz\Kstpm}}\xspace}
\newcommand{\ymbdkst}{\ensuremath{y_{-}^{\Dz\Kstpm}}\xspace}
\newcommand{\xpbdkst}{\ensuremath{x_{+}^{\Dz\Kstpm}}\xspace}
\newcommand{\ypbdkst}{\ensuremath{y_{+}^{\Dz\Kstpm}}\xspace}
\newcommand{\xixbdstk}{\ensuremath{x_{\xi}^{\Dstz\Kpm}}\xspace}
\newcommand{\xiybdstk}{\ensuremath{y_{\xi}^{\Dstz\Kpm}}\xspace}
\newcommand{\xmbdstk}{\ensuremath{x_{-}^{\Dstz\Kpm}}\xspace}
\newcommand{\ymbdstk}{\ensuremath{y_{-}^{\Dstz\Kpm}}\xspace}
\newcommand{\xpbdstk}{\ensuremath{x_{+}^{\Dstz\Kpm}}\xspace}
\newcommand{\ypbdstk}{\ensuremath{y_{+}^{\Dstz\Kpm}}\xspace}

\section{Sensitivity studies}

Monte Carlo simulation is used to evaluate the impact of applying a simultaneous
measurement technique on the sensitivity to $\zpm$, and subsequently the weak phase $\g$, using the BPGGSZ approach.
A simplified model description of the $D$ meson decay amplitude over its phase space is implemented~\cite{BaBarDModel} in order
to generate events. This has been cross-checked with a slightly more sophisticated model~\cite{LHCbgammaGGSZDK,cfit} and gives
identical results.
Two thousand signal-only pseudo-experiments are generated,
with each pseudo-experiment including five $B$
meson decay modes $\bdk$, $\bzdkst$, $\bdstk$, $\bdkst$ and $\bdpi$, where the $D$ decays into the final state $K_S^0\pi^+\pi^-$.
Experimental effects such as variation of the efficiency across the phase space, variation in the amplitude model, background contributions and momentum resolution have not been considered in this paper. It is expected that inclusion of these effects, incorporating cross-feed and correlations between decay modes, will have a small impact on the overall conclusions drawn by this study. 
Ensembles of pseudo-experiments are generated with sample sizes equivalent to $3$~fb$^{-1}$ and $9$~fb$^{-1}$ of LHCb data, representing the milestones
reached after Run 1 and Run 2 of the LHC, respectively.
World average values~\cite{HFLAV18,PDG2018} are used for the hadronic parameters $r_m$ and $\delta_m$ in each
$B$ meson decay mode, with $\g=70\degrees$ in all cases.
Since current experimental constraints on the admixture coefficients for $\bdpi$
decays are rather loose~\cite{LHCb:GammaCombo2016},
values of $r_{D\pi} = 0.005$ and $\delta_{D\pi} = 300\degrees$ are used in the \bdpi mode.

\newcommand{\floatkappa}{\mcolr{1}{float $\kappa$}}
\newcommand{\fixkappa}  {\mcolr{1}{fix   $\kappa$}}

\newcommand{\dxm}    {\delta x_-}
\newcommand{\dym}    {\delta y_-}
\newcommand{\dxp}    {\delta x_+}
\newcommand{\dyp}    {\delta y_+}
\newcommand{\drxDkstz}{{\delta \re{\xi_{DK^{\star0}}}}}
\newcommand{\dixDkstz}{{\delta \im{\xi_{DK^{\star0}}}}}
\newcommand{\drxDkst}{{\delta \re{\xi_{DK^{\star}}}}}
\newcommand{\dixDkst}{{\delta \im{\xi_{DK^{\star}}}}}
\newcommand{\drxDstk}{{\delta \re{\xi_{D^{\star}K}}}}
\newcommand{\dixDstk}{{\delta \im{\xi_{D^{\star}K}}}}
\newcommand{\drxDpi} {{\delta \re{\xi_{D\pi}}}}
\newcommand{\dixDpi} {{\delta \im{\xi_{D\pi}}}}
\newcommand{\sxm}    {\sigma_{x_-}}
\newcommand{\sym}    {\sigma_{y_-}}
\newcommand{\sxp}    {\sigma_{x_+}}
\newcommand{\syp}    {\sigma_{y_+}}
\newcommand{\sxmmode}    {\sigma_{x_-^m}}
\newcommand{\symmode}    {\sigma_{y_-^m}}
\newcommand{\sxpmode}    {\sigma_{x_+^m}}
\newcommand{\sypmode}    {\sigma_{y_+^m}}
\newcommand{\srxDkstz}{{\sigma_{\re{\xi_{DK^{\star0}}}}}}
\newcommand{\sixDkstz}{{\sigma_{\im{\xi_{DK^{\star0}}}}}}
\newcommand{\srxDkst}{{\sigma_{\re{\xi_{DK^{\star}}}}}}
\newcommand{\sixDkst}{{\sigma_{\im{\xi_{DK^{\star}}}}}}
\newcommand{\srxDstk}{{\sigma_{\re{\xi_{D^\star K}}}}}
\newcommand{\sixDstk}{{\sigma_{\im{\xi_{D^\star K}}}}}
\newcommand{\srxDpi} {{\sigma_{\re{\xi_{D\pi}}}}}
\newcommand{\sixDpi} {{\sigma_{\im{\xi_{D\pi}}}}}
\newcommand{\blank}  {\mcolr{1}{\mbox{---}}}

\newcommand{\tr}[1]{\mrow{2}{$#1$}}
\newcommand{\trblank}{\mrow{2}{\mbox{---}}}

A standalone fit to each pseudo-experiment with each separate decay mode is used as a reference. In this case each decay mode
is used to independently determine $z^{m}_\pm$, resulting in a total of $4N$ parameters.
Several simultaneous fits are performed, progressively adding each decay mode to the set of decays considered.
In this case, $z_\pm = x_\pm + i\,y_\pm$ and the appropriate $\xi_m$ parameters are determined, resulting in a total of $2(N+1)$ parameters.
The obtained statistical uncertainties on these parameters, \ie~the width of the distribution of fitted values across the ensemble,
 assuming $3$~fb$^{-1}$ of LHCb data, are summarised in tables~\ref{tb.results.sim} and~\ref{tb.results.stand}.
 For all of these parameters, when performing
the simultaneous fit, the results exhibit unbiased Gaussian behaviour and the uncertainty estimates provide the appropriate coverage.
This is not the case for all parameters when performing the standalone fits,
particularly in the \bdpi and \bdkst decay modes. In these cases, when fitting the standalone parameters $z^{m}_\pm$ (for $m=D\pipm$ or $m=D\Kstarpm$), the results do not demonstrate Gaussian behaviour, with biases as large as 40\% of the statistical uncertainty.
The reason for this in the \bdkst mode is that the expected number of events is very low ($\sim 90$ at 3\invfb), therefore the $D$ decay phase space is very
sparsely populated and these fits are far from the Gaussian regime. In the \bdpi case, where $r_m$ is very small,
the Cartesian parameters $z^m_\pm$ are very close to zero, which impacts on fit stability and reliable error estimation of the polar parameters.
Both of these issues are resolved when reparameterising the problem in terms of $\xi$ using the simultaneous fit methodology outlined here.

\newcommand{\mcl}[1]{\multicolumn{1}{l}{#1}}
\newcommand{\mcr}[1]{\multicolumn{1}{r}{#1}}

\begin{table}
\renewcommand{\tabcolsep}{2mm}
\renewcommand{\arraystretch}{1.3}
{\fontsize{10pt}{11.4pt}\selectfont
\begin{tabular}{ p{2cm}  c c c c c c c c c c c c}
\toprule
            &  \mcl{\bdk}         &  \mcl{\bdk,}  &  \mcl{\bdk,}   &  \mcl{\bdk,}    &  \mcl{\bdk,}    \\
            &                     &  \mcl{\bdpi}  &  \mcl{\bdpi,}  &  \mcl{\bdpi,}   &  \mcl{\bdpi,}   \\
 Channel(s) &                     &               &  \mcl{\bzdkst} &  \mcl{\bzdkst,} &  \mcl{\bzdkst,} \\
            &                     &               &                &  \mcl{\bdstk}   &  \mcl{\bdstk,}  \\
            &                     &               &                &                 &  \mcl{\bdkst}   \\
\midrule
 $\sxm     $     & $0.0193 \pm 0.0003$ & $0.0199 \pm 0.0003$ & $0.0197 \pm 0.0003$ & $0.0191 \pm 0.0003$ & $0.0192 \pm 0.0003$ \\
 $\sym     $     & $0.0220 \pm 0.0004$ & $0.0222 \pm 0.0004$ & $0.0214 \pm 0.0003$ & $0.0206 \pm 0.0003$ & $0.0210 \pm 0.0004$ \\
 $\sxp     $     & $0.0200 \pm 0.0003$ & $0.0202 \pm 0.0003$ & $0.0195 \pm 0.0003$ & $0.0192 \pm 0.0003$ & $0.0194 \pm 0.0003$ \\
 $\syp     $     & $0.0214 \pm 0.0004$ & $0.0218 \pm 0.0004$ & $0.0210 \pm 0.0003$ & $0.0204 \pm 0.0003$ & $0.0209 \pm 0.0004$ \\
 $\srxDpi  $     & $        -        $ & $0.0420 \pm 0.0008$ & $0.0418 \pm 0.0007$ & $0.0419 \pm 0.0008$ & $0.0419 \pm 0.0007$ \\
 $\sixDpi  $     & $        -        $ & $0.0392 \pm 0.0007$ & $0.0395 \pm 0.0007$ & $0.0392 \pm 0.0007$ & $0.0394 \pm 0.0007$ \\
 $\srxDkstz$     & $        -        $ & $        -        $ & $0.9011 \pm 0.0176$ & $0.9045 \pm 0.0176$ & $0.9003 \pm 0.0176$ \\
 $\sixDkstz$     & $        -        $ & $        -        $ & $0.8557 \pm 0.0161$ & $0.8571 \pm 0.0160$ & $0.8562 \pm 0.0162$ \\
 $\srxDstk $     & $        -        $ & $        -        $ & $        -        $ & $0.5189 \pm 0.0090$ & $0.5173 \pm 0.0087$ \\
 $\sixDstk $     & $        -        $ & $        -        $ & $        -        $ & $0.5351 \pm 0.0098$ & $0.5356 \pm 0.0100$ \\
 $\srxDkst $     & $        -        $ & $        -        $ & $        -        $ & $        -        $ & $0.6847 \pm 0.0116$ \\
 $\sixDkst $     & $        -        $ & $        -        $ & $        -        $ & $        -        $ & $0.7115 \pm 0.0128$ \\
\bottomrule
\end{tabular}
}
\caption{Expected statistical uncertainties of the cartesian parameters when fitting using the simultaneous method, adding successive decay modes, assuming
	$3$~fb$^{-1}$ of LHCb data.}
  \label{tb.results.sim}
\end{table}

\begin{table}
\renewcommand{\tabcolsep}{2.5mm}
\renewcommand{\arraystretch}{1.3}
{\fontsize{10pt}{11.4pt}\selectfont
\begin{tabular}{ p{1.9cm} c c c c c c c c c c c c}
\toprule
Channel &  \mcl{\bdk}         &  \mcl{\bdpi}         &  \mcl{\bzdkst}         &  \mcl{\bdstk}         &  \mcl{\bdkst}         \\
\midrule
 $\sxmmode     $     & $0.0193 \pm 0.0003$ & $0.0050 \pm 0.0001$ & $0.1112 \pm 0.0019$ & $0.0636 \pm 0.0011$ & $0.0900 \pm 0.0015$ \\
 $\symmode     $     & $0.0220 \pm 0.0004$ & $0.0056 \pm 0.0001$ & $0.1134 \pm 0.0020$ & $0.0668 \pm 0.0012$ & $0.1017 \pm 0.0018$ \\
 $\sxpmode     $     & $0.0200 \pm 0.0003$ & $0.0054 \pm 0.0001$ & $0.1064 \pm 0.0020$ & $0.0687 \pm 0.0013$ & $0.0841 \pm 0.0014$ \\
 $\sypmode     $     & $0.0214 \pm 0.0004$ & $0.0055 \pm 0.0001$ & $0.1112 \pm 0.0021$ & $0.0769 \pm 0.0014$ & $0.0911 \pm 0.0015$ \\
 \bottomrule
\end{tabular}
}
	\caption{Expected statistical uncertainties of the cartesian parameters when fitting each decay mode with the standalone fit method, assuming
	$3$~fb$^{-1}$ of LHCb data.}
  \label{tb.results.stand}
\end{table}

Using the pseudo-experiment sizes considered in this study, the inclusion of $\bdpi$ decays
does not improve the sensitivity to $\gamma$, mainly because the size of the interference in this mode is very small ($r_{D\pi}\approx 0.005$). However, inclusion of this high
statistics mode is very important when fitting experimental data because it can be used to determine the variation in candidate reconstruction and selection efficiency across the $D \rightarrow K_S^0\pi^+\pi^-$phase space~\cite{LHCbgammaGGSZDK}.
These studies show that inclusion of this decay mode is statistically robust when using the simultaneous parameterisation, which is not the case when treating it as an independent mode.
Furthermore, the addition of each other
decay mode contributes to a smaller uncertainty on $\zpm$.

\begin{figure}
\vspace{0.5cm}
  \centering
  \includegraphics[width=0.48\textwidth]{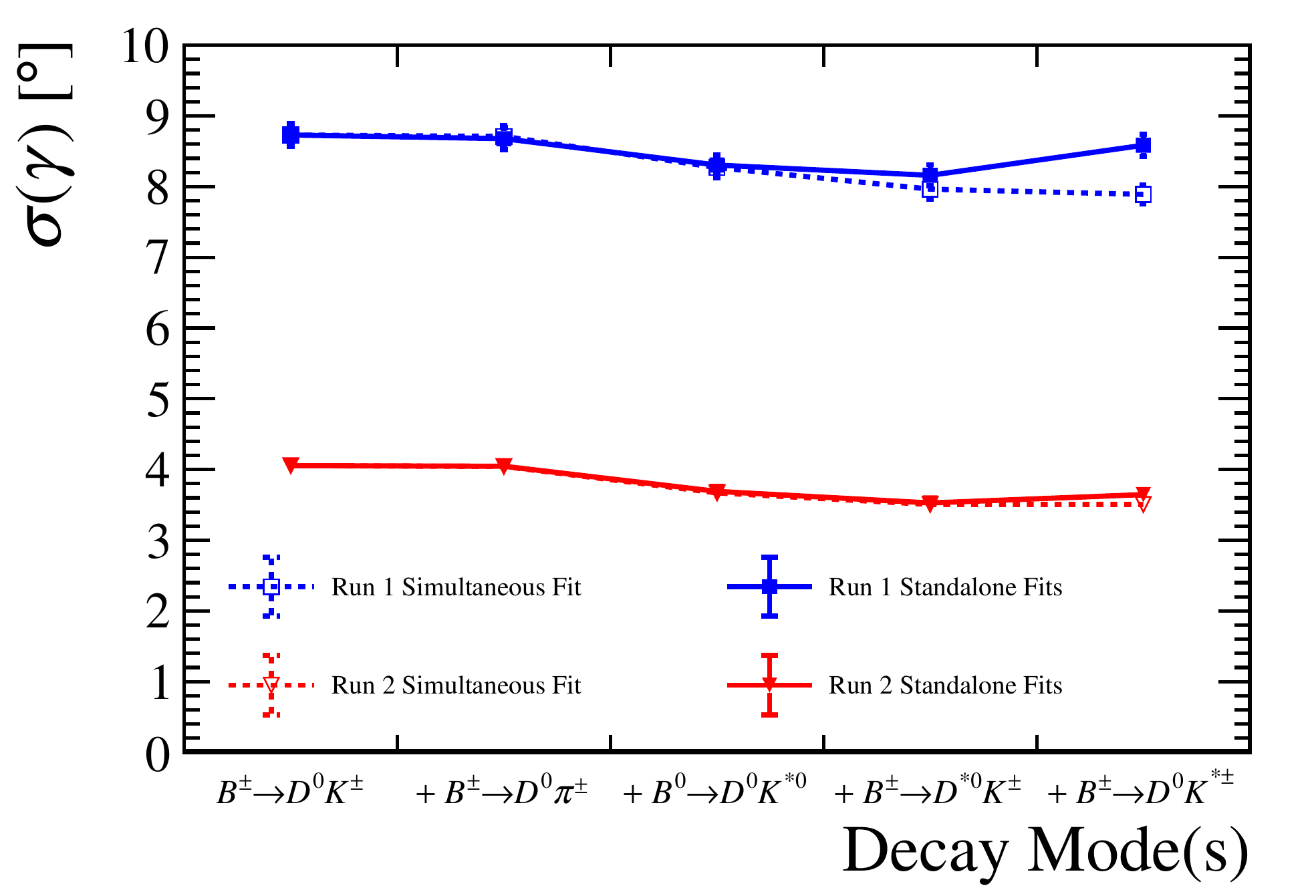}
	\caption{Progression of the expected statistical uncertainty on CKM angle $\gamma$ when incorporating additional decay modes using the simultaneous approach (dashed lines)
	and combining the standalone measurements (solid lines) with pseudo-experiment sample sizes corresponding to Run 1 ($3$~fb$^{-1}$, blue) and Run 2 ($9$~fb$^{-1}$, red) of LHCb data.}
  \label{fg.gamma_err}
\end{figure}

After the simultaneous and standalone fits above have been performed, the fitted values and full covariance matrix for each pseudo-experiment
are used to determine a value for the \CP-violating weak phase \g.
The corresponding results for the expected sensitivity
when progressively adding
decay modes are shown in table~\ref{tb.results.gamma} and figure~\ref{fg.gamma_err}.
It can be seen that the simultaneous method provides a marginal gain in sensitivity over the standalone method, although
its main advantage, as highlighted above, is its statistical robustness. Moreover, when determining the sensitivity to \g using the standalone method,
the pull distributions for the hadronic ratio parameters, $r_{m}$, show large biases and widths far from unity. Given that the uncertainty on \g
is inversely proportional to the central value of $r_{m}$ this results in inaccurate estimations of the uncertainty on \g when using the standalone method. Consequently,
the quoted values for the expected precision on \g from the standalone fits actually fluctuate around their unknown true values, and should be interpreted accordingly. This can cause behaviour such as the uncertainty appearing to increase when adding more measurements, which occurs in figure~\ref{fg.gamma_err}, although it is typically mitigated
as the sample sizes increase.

\begin{figure}
\vspace{0.5cm}
  \centering
  \includegraphics[width=0.48\textwidth]{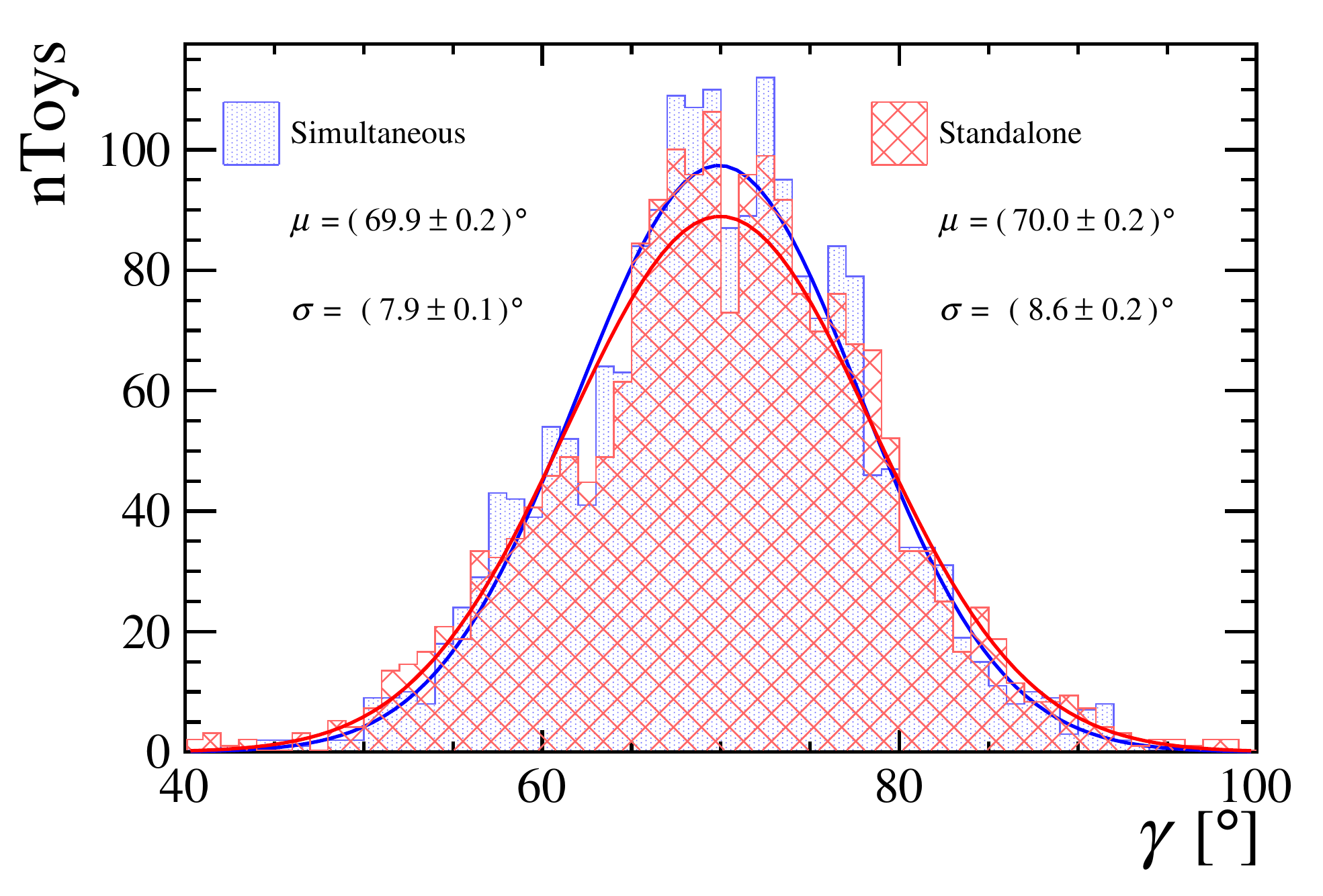}
  \includegraphics[width=0.48\textwidth]{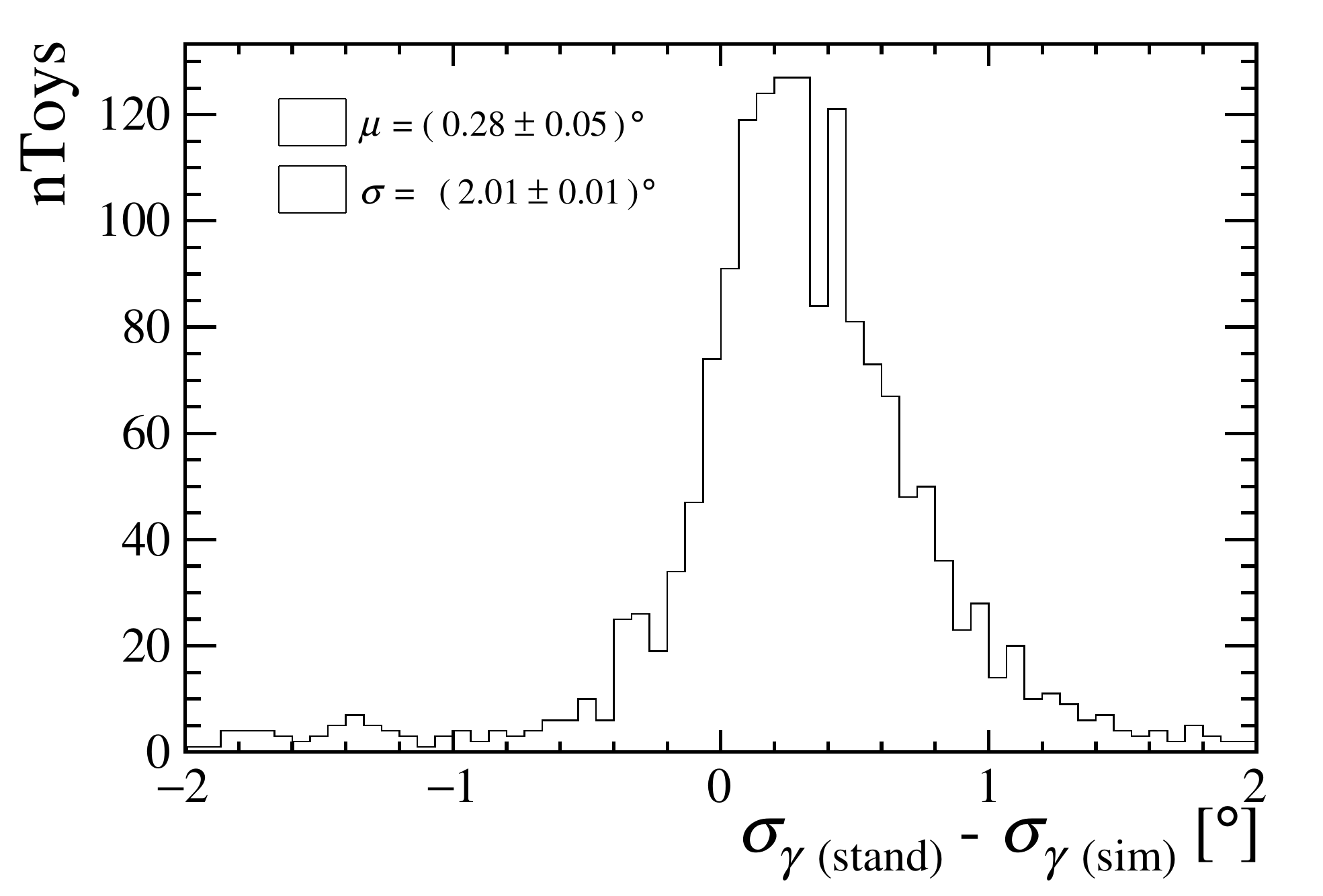}
  \caption{Left: Comparison between the simultaneous approach (blue) and combination of standalone measurements (red) for the value
    of $\gamma$ determined with pseudo-experiments corresponding to a sample size of 3\invfb of LHCb data. The solid lines show a Gaussian fit
    to each distribution, with the mean and width of the fitted Gaussian shown
    in the top left and top right, respectively. Right: The difference of the fitted uncertainty on the CKM angle \g obtained from pseudo-experiments when
		using the simultaneous approach and a combination
		of standalone measurements.     }
  \label{fg.gamma_sensitivity}
\end{figure}

Figure~\ref{fg.gamma_sensitivity} (left) shows a comparison between the distribution of fitted \g values for the two methods when all five decay modes are included and a sample size corresponding to 3\invfb of LHCb data is used. In a small fraction of cases for the standalone pseudo-experiments ($\sim3\%$) the fit fails converge whereas for the simultaneous extraction this has not been found to happen. The distributions shown in figure~\ref{fg.gamma_sensitivity} (left) have $0.1\%$ (3.5\%) of events in underflow or overflow bins for the simultaneous (standalone) extractions. This demonstrates that the rate of outliers seems to be much higher in the standalone case.
The right-hand figure shows, per pseudo-experiment, the difference between the uncertainty on \g determined from the standalone method and
from the simultaneous method.
On average, there is a small ($0.3^\circ$) gain when fitting simultaneously, but the difference can be as
large as $2-3^\circ$ in either direction.
However, as previously stated, the evaluated uncertainties for the standalone method are less reliable for the \bdpi and \bdkst decay modes. This is highlighted by further analysis of the far reaching tails in figure~\ref{fg.gamma_sensitivity} which reveals that for events with $\sigma_{\gamma\mathrm{(stand)}}-\sigma_{\gamma\mathrm{(sim)}}<0.5 \degrees$ or $\sigma_{\gamma\mathrm{(stand)}}-\sigma_{\gamma\mathrm{(sim)}}>1 \degrees$ the width of the distribution of values for $\gamma$ is $8.5\degrees$ ($14\degrees)$ for the simultaneous (standalone) extraction. Furthermore, for these events the normalised pull distributions have a mean of $-0.03\pm0.05$ ($-0.05\pm0.06$) and a width of $1.09\pm0.04$ ($1.40\pm0.04$) for the simultaneous (standalone) extraction demonstrating the less reliable error estimation in the standalone case.

It should be noted that the statistical precision on \g obtained in any future experimental analyses employing this technique will depend on the exact results obtained for $\zpm$ and $\xi_m$ from data. Nevertheless,
the technique provides a rigorous and straightforward treatment of the uncertainties,
allowing for simple inclusion of correlations between various decay modes. The studies presented here demonstrate good
statistical behaviour and a modest improvement in sensitivity when using the reparameterisation of equation~\ref{eq.xicdefn} in signal-only simulations. It may be that the
improvement is larger when also including experimental backgrounds, efficiency variations and systematic effects.

\begin{table}
\renewcommand{\tabcolsep}{2mm}
\renewcommand{\arraystretch}{1.3}
{\fontsize{10pt}{11.4pt}\selectfont
\begin{tabular}{l  l  c c c c c c c c c c c c}
\toprule
                 & &  \mcl{\bdk}         &  \mcl{\bdk,}  &  \mcl{\bdk,}   &  \mcl{\bdk,}    &  \mcl{\bdk,}    \\
                 & &                     &  \mcl{\bdpi}  &  \mcl{\bdpi,}  &  \mcl{\bdpi,}   &  \mcl{\bdpi,}   \\
\multicolumn{2}{l}{Channel(s)} &                     &               &  \mcl{\bzdkst} &  \mcl{\bzdkst,} &  \mcl{\bzdkst,} \\
                 & &                     &               &                &  \mcl{\bdstk}   &  \mcl{\bdstk,}  \\
                 & &                     &               &                &                 &  \mcl{\bdkst}   \\
\midrule
	\multirow{2}{*}{Simultaneous} & Run 1 & $(8.73 \pm 0.16)^\circ$ & $(8.71 \pm 0.15)^\circ$ & $(8.27 \pm 0.15)^\circ$ & $(7.97 \pm 0.15)^\circ$ & $(7.89 \pm 0.13)^\circ$ \\
	& Run 2 & $(4.05 \pm 0.07)^\circ$ & $(4.04 \pm 0.08)^\circ$ & $(3.67 \pm 0.06)^\circ$ & $(3.51 \pm 0.06)^\circ$ & $(3.50 \pm 0.06)^\circ$ \\
\midrule
	\multirow{2}{*}{Standalone}   & Run 1 & $(8.73 \pm 0.16)^\circ$ & $(8.68 \pm 0.15)^\circ$ & $(8.31 \pm 0.15)^\circ$ & $(8.16 \pm 0.15)^\circ$ & $(8.47 \pm 0.13)^\circ$ \\
	& Run 2 & $(4.05 \pm 0.07)^\circ$ & $(4.05 \pm 0.08)^\circ$ & $(3.69 \pm 0.06)^\circ$ & $(3.53 \pm 0.06)^\circ$ & $(3.64 \pm 0.07)^\circ$ \\
\bottomrule

\end{tabular}
}
\caption{Expected statistical uncertainty on CKM angle $\gamma$ from the simultaneous (top part) and standalone (bottom part) fit methods when incorporating additional decay modes.}
  \label{tb.results.gamma}
\end{table}

\section{Conclusions}

An approach to simultaneously measuring the \CP parameters $z_\pm$ sensitive to the CKM angle $\gamma$
in multiple $B$ meson decays has been presented.
The formalism reduces the number of free parameters,
allows for the consideration of experimentally reconstructed decays that are signal in one case but background in another,
and allows for a common treatment of systematic uncertainties.
Sensitivity studies show that including additional $B$ meson decay modes
contributes marginally to a smaller uncertainty on $\zpm$ but has considerably safer statistical behaviour,
and thus has the potential to offer an improvement
compared to a combination of standalone measurements.
This is likely to be further enhanced when systematic uncertainties and their correlations are also considered,
particularly for large future data sets.

\section{Acknowledgements}

The authors wish to thank their colleagues on the LHCb experiment for the fruitful and enjoyable collaboration that inspired this study.
In particular, they would like to thank Tim Gershon for helpful comments on the manuscript.
This work was supported by the Science and Technology Facilities Council (STFC). MK is supported by the STFC under grant \#ST/R004536/1.

\addcontentsline{toc}{section}{References}
\bibliographystyle{LHCb}
\bibliography{biblio}

\end{document}